\documentclass{aa}
\usepackage{graphics}
\begin{document}

\title{The extended atmosphere and evolution of the RV~Tau star, R~Scuti
       \thanks{
       Based on observations with ISO, an ESA project with instruments
       funded by ESA Member States (especially the PI countries: France,
       Germany, the Netherlands and the United Kingdom) with the
       participation of ISAS and NASA. The SWS is a joint project of
       SRON and MPE.}
}

\titlerunning{The atmosphere and evolution of  R~Sct}
\authorrunning{M.~Matsuura et al.}

\offprints{M.M. (m.matsuura@umist.ac.uk)}
\date{Received: 03 December 2001; Accepted: 11 March 2002}

\author{ M.~Matsuura       \inst{1}
    \and  I.~Yamamura       \inst{2}
    \and  A.A.~Zijlstra     \inst{1}
    \and  T.R.~Bedding        \inst{3}
}
\institute{
       Department of Physics, UMIST, P.O. Box 88, Manchester M60 1QD, UK
    \and
       Institute of Space and Astronautical Science (ISAS),
       Yoshino-dai 3-1-1, Sagamihara, Kanagawa, 229-8510, Japan
    \and
       School of Physics, University of Sydney 2006, Australia
}

\abstract{ We analyze ISO/SWS spectra of the RV~Tau star R~Scuti.  
The infrared spectra are dominated by H$_2$O emission bands.  The near-
and mid-infrared excess is attributed to H$_2$O; the dust contribution is
less important.  We also identify CO, SiO and CO$_2$ bands.  The
various molecular emission bands originate from an extended
atmosphere, an atmosphere above the photosphere.
The extended atmosphere of R~Sct
is formed from matter which gradually have lifted up from
the photosphere through the pulsations of the star.
In contrast to the
abundant molecules around the star, the silicate dust feature is weak
and the dust mass-loss rate is only 
$\dot M_{\mathrm d}=10^{-11}$~M$_{\sun}$\,yr$^{-1}$.  
This implies that there might be
a process to inhibit dust formation from molecules.  
RV Tau stars are commonly considered as post-AGB stars. 
While a detached dust envelope around R Sct 
is consistent with such an
interpretation, we show that its period evolution is slower than
expected.  We argue that R Sct may be a thermal-pulsing AGB star,
observed in a helium-burning phase.
%
}

\maketitle

\section{Introduction}
\object{RV~Tau stars} are pulsating variables characterized by
alternating deep and shallow minima in their light curves.  They are
generally considered to be \object{post-AGB stars} with low initial
masses ($\sim$1\,$M_{\sun}$; Jura~\cite{Jura86}).  The abundance
ratios show that they have experienced first dredge-up at the bottom of the
red giant branch (Giridhar et al.\ \cite{Giridhar00}).
Based on their infrared dust excesses, RV~Tau stars are classified
into two groups: those with extensive warm dust and those without
evidence of dust in the near-infrared region (Goldsmith et al.\
\cite{Goldsmith87}). \object{R~Sct} is the brightest star in the
visible in the latter group.

R~Sct (HR 7066; HIP 92202) has a reported period of 147~days (Kholopov
et al.\ \cite{Kholopov88}).  The effective temperature varies from
4750 to 5250~K (Shenton et al.\ \cite{Shenton94}); the spectral type may
vary as late as M3 at minimum phase (Kholopov et al.\ \cite{Kholopov88}). 

An atmosphere beyond the photosphere,
which we call an extended atmosphere later on in this paper,
has been found in R~Sct from
the analysis of sodium lines (L\`{e}bre \& Gillet~\cite{Lebre91}).
High-resolution spectroscopic observations 
have uncovered CO, H$_2$O
and OH molecules in the K-band (Mozurkewich et al.\ \cite{Mozurkewich87}).  
The complicated variation of the velocity
profile shows that the CO originates in the upper atmosphere, above
the photosphere, which is affected by the pulsation.  Since the
spectral-type of R~Sct is too early for the presence of H$_2$O, they
conclude that the observed H$_2$O is circumstellar origin.

In this paper, we analyze spectra of R~Sct taken with 
the Short-Wavelength Spectrometer (SWS; de~Graauw et al.\ \cite{deGraauw96}) on
board the Infrared Space Observatory (ISO; Kessler et al.\
\cite{Kessler96}).  Because there is no interference from molecules in
the terrestrial atmosphere, ISO spectra are ideal for the study of
molecules in stars.  We find that the IR spectra of R~Sct is
dominated by molecular emission features, especially from H$_2$O.
In addition, SiO and CO$_2$ bands are identified.  We discuss the
distribution of these molecules and the mass-loss history. The 200-year
light curve is used to address the evolution of R Sct.

\section{Observational data and analysis}

\begin{figure}
   \resizebox{\hsize}{!}{\includegraphics*[75,270][500,555]{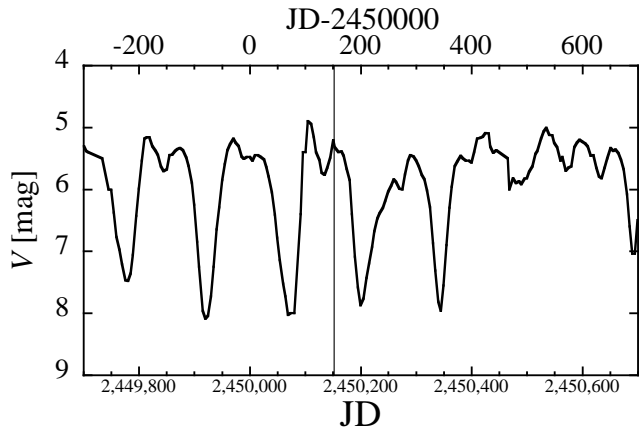}}
\caption{
The visual light curve of R~Sct (Mattei~\cite{Mattei95},
private communication).
The magnitudes are averaged in 5-day bins.
The date at which the ISO/SWS observed this star
is shown by the vertical line.}
    \label{aavso}
\end{figure}

  We obtained the SWS data for R~Sct from the ISO Data Archive.  R~Sct
was observed on 10 March, 1996 by programme of MOLBANDS (P.I. A. Heske).  
The phase is estimated as $\phi$=0.60
from the AAVSO light curve (Fig.~\ref{aavso}; Mattei~\cite{Mattei95},
private communication), where $\phi=0.0$ corresponds to the nearest
previous deep minimum.  The spectra were obtained using the
full-grating scan mode (AOT~01, scan speed 1).  The wavelength range
covers 2.35--45.2~$\mu$m, but the signal-to-noise ratio is poor above
20~$\mu$m and these data are not used.  The spectral resolution
is $\lambda / \Delta \lambda =300$--500, depending on the wavelength.
The data were reduced using the SWS Interactive Analysis package.  The
calibration parameters of October 1999 were used for the wavelength,
detector responsivity, and absolute flux calibrations.  The spectra
were re-gridded to a constant wavelength resolution of $\lambda /
\Delta \lambda =300$.

  To identify the molecular features, we calculate molecular spectra
using a circular slab model, which was used previously for analysing
spectra of \object{Mira variables} (e.g., Yamamura et al.\
\cite{Yamamura99}).  We adopt a multiple layer model.
This is an approximation for a shell that shows a
temperature and density gradient; the model layers do not necessarily
represent discrete shells.  Line lists for H$_2$O, CO$_2$, SiO and CO
were taken from Partridge \& Schwenke (\cite{Partridge97}), HITRAN
(Rothman et al.\ \cite{Rothman01}), Langhoff \& Bauschlicher
(\cite{Langhoff93}), and HITEMP (Rothman et al.\ \cite{Rothman01b}),
respectively.  The solar isotopic abundance ratio is adopted for
H$_2$O and SiO.  Bujarrabal et al.\ (\cite{Bujarrabal90}) derive
$^{12}$C/$^{13}$C $\la$10 and we assume 10~\% of carbon atoms 
are $^{13}$C.
For the energy level populations of the molecules, local thermodynamic
equilibrium (LTE) is assumed.  A line width of 5~km\,s$^{-1}$
is assumed for all the molecules.  We use a black body as the
background source illuminating the molecular layers. 
We assume that the background is a black-body with a temperature of 5000~K.
The near- and mid-infrared spectrum, which is in the Rayleigh Jeans region of 
the black body, is insensitive to the black-body temperature.
The synthesized 
spectrum is not significantly affected by the black-body temperature
with in the range 5000--7000\,K.

\section{Results}
  
\begin{figure*}
   \resizebox{\hsize}{!}{\includegraphics*[30,280][565,525]{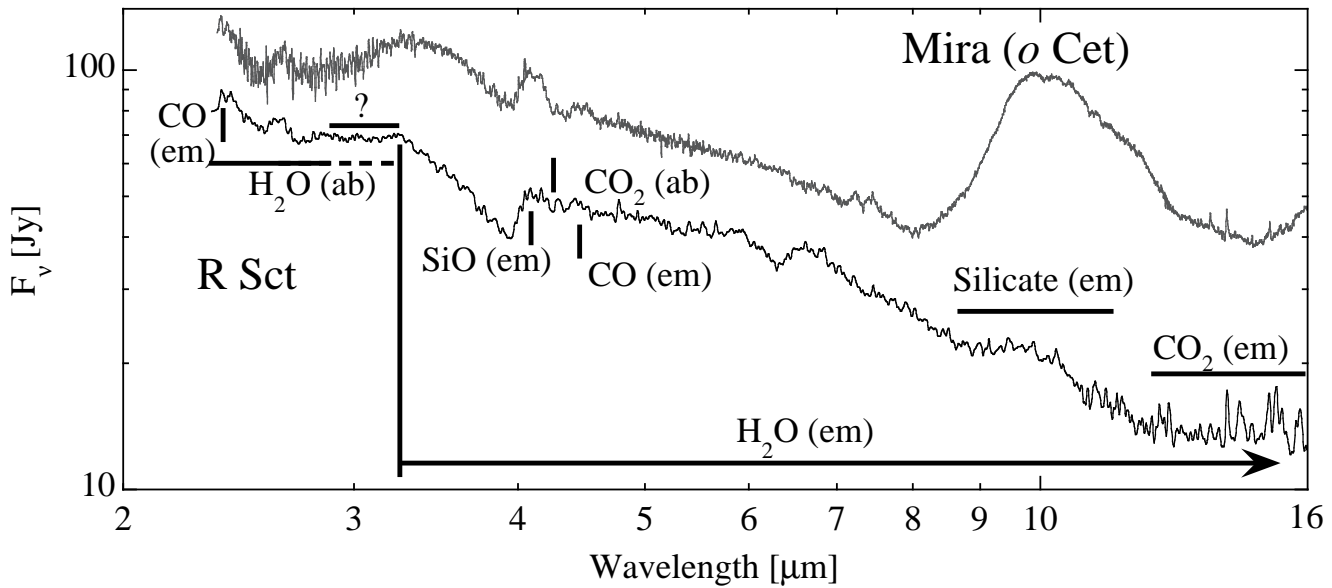}}
\caption{
The ISO/SWS spectra of R Sct (black line).
The locations of the major molecular features on R~Sct are indicated.
The spectra of R~Sct  is compared with that of
an oxygen-rich AGB star, $o$~Cet (grey line; flux is scaled).
These two stars show similar spectra below 4.5~$\mu$m.
Note that the spectral resolution of $o$~Cet is
twice as high as that of R~Sct.
}
    \label{rsct_ocet}
\end{figure*}

The ISO/SWS spectra of R~Sct are shown in Fig.~\ref{rsct_ocet}.  Below
4.5~$\mu$m, the spectra are similar to those of a typical AGB
star, Mira (\object{$o$~Cet}; spectral type of M5e--M9e), which is dominated
by water-vapour emission (Yamamura et al.\ \cite{Yamamura99}). 
Water-vapour bands are mostly
seen in emission in the near- and mid-IR spectra of R~Sct, except at
$\sim$2.7~$\mu$m where it is in absorption.  CO$_2$ is seen in emission at
13--16~$\mu$m and in absorption at 4.2~$\mu$m.  CO at 2.3~$\mu$m and
at 4.6~$\mu$m is seen in emission.  SiO emission features are detected at
4.1~$\mu$m.  A weak silicate dust excess is present around 10~$\mu$m.

\begin{figure*}
   \resizebox{\hsize}{!}{\includegraphics*[30,250][560,565]{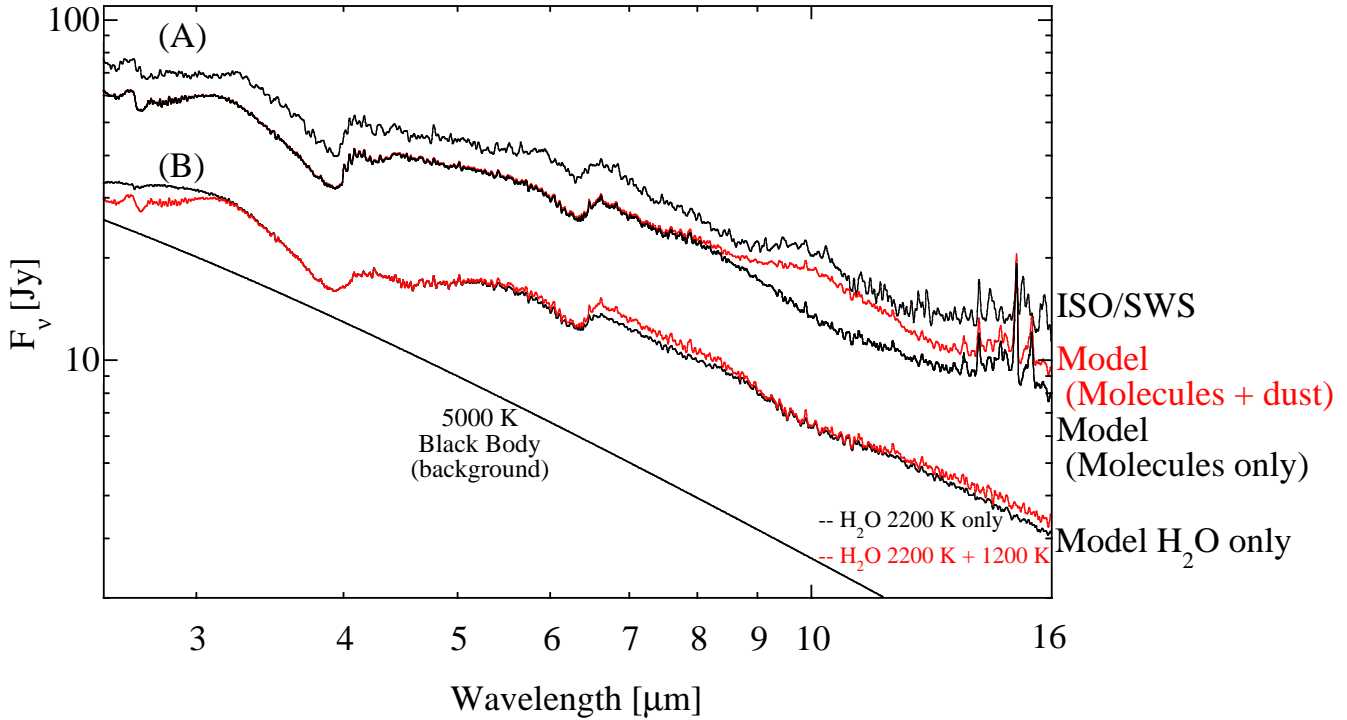}}
\caption{ (A) The ISO/SWS spectra of R~Sct (top) are compared with synthesized
spectra, which are offset.  One
synthesized spectra includes molecules only; the other
also includes a dust excess.  
Most part of the ISO/SWS spectra are well reproduced by the
synthesized spectra containing only the molecular bands.  A weak dust
excess is observed at 10~$\mu$m.
(B) Synthesized spectra containing only H$_2$O.  The black line shows
the 2200\,K H$_2$O layer only, and the grey line shows both H$_2$O
layers (2200~K and 1200~K).  For comparison, the background source, a
5000~K black body is also presented.  H$_2$O bands are seen in
emission; the global shape of the spectra are dominated by H$_2$O
bands with an excitation temperature of 2200\,K.  The 1200\,K component
contributes to the absorption at 2.7~$\mu$m and the emission at
6.2~$\mu$m.  } \label{rsct_model}
\end{figure*}
\begin{table}
\begin{caption}
{ The parameters of synthesized molecular spectra.  We adopt two
H$_2$O components: one is needed to reproduce the absorption at
2.7~$\mu$m and the other causes the strong emission features over the
whole wavelength region. The size of the water layer, $R$, is given
relative to the stellar radius ($R_*$).  The features are indicated as
either emission (em) or absorption (ab).  }
\label{table-parameters}
\end{caption}
\begin{tabular}{l rrr l} \hline
  & $T_{\mathrm ex}$  & $N$  & $R$  & Appearance \\
  & [K] & [cm$^{-2}$] & [$R_*$] &  \\ \hline
CO & 4000 & $3.0\times10^{21}$ & 1.6 & em \\ 
H$_2$O-1   & 2200 & $1.2\times10^{21}$ & 2.8 &em\\
SiO      & = $T_{\mathrm{H_2O}-1}$ & $8.0\times10^{20}$  & =$R_{\mathrm{H_2O}-1}$ &em 
\\ 
H$_2$O-2   & 1200 & $5.0\times10^{19}$ & 5.5 & ab (2.7~$\mu$m)\\
  & & & & em (6.2~$\mu$m)\\
CO$_2$   & 800 &$3\times10^{18}$   & 10.5 & ab 
(4.2~$\mu$m) \\
  & & & & em (13~$\mu$m) \\ \hline
\end{tabular} \\
\end{table}

\begin{figure}
   \resizebox{\hsize}{!}{\includegraphics*[85,130][570,830]{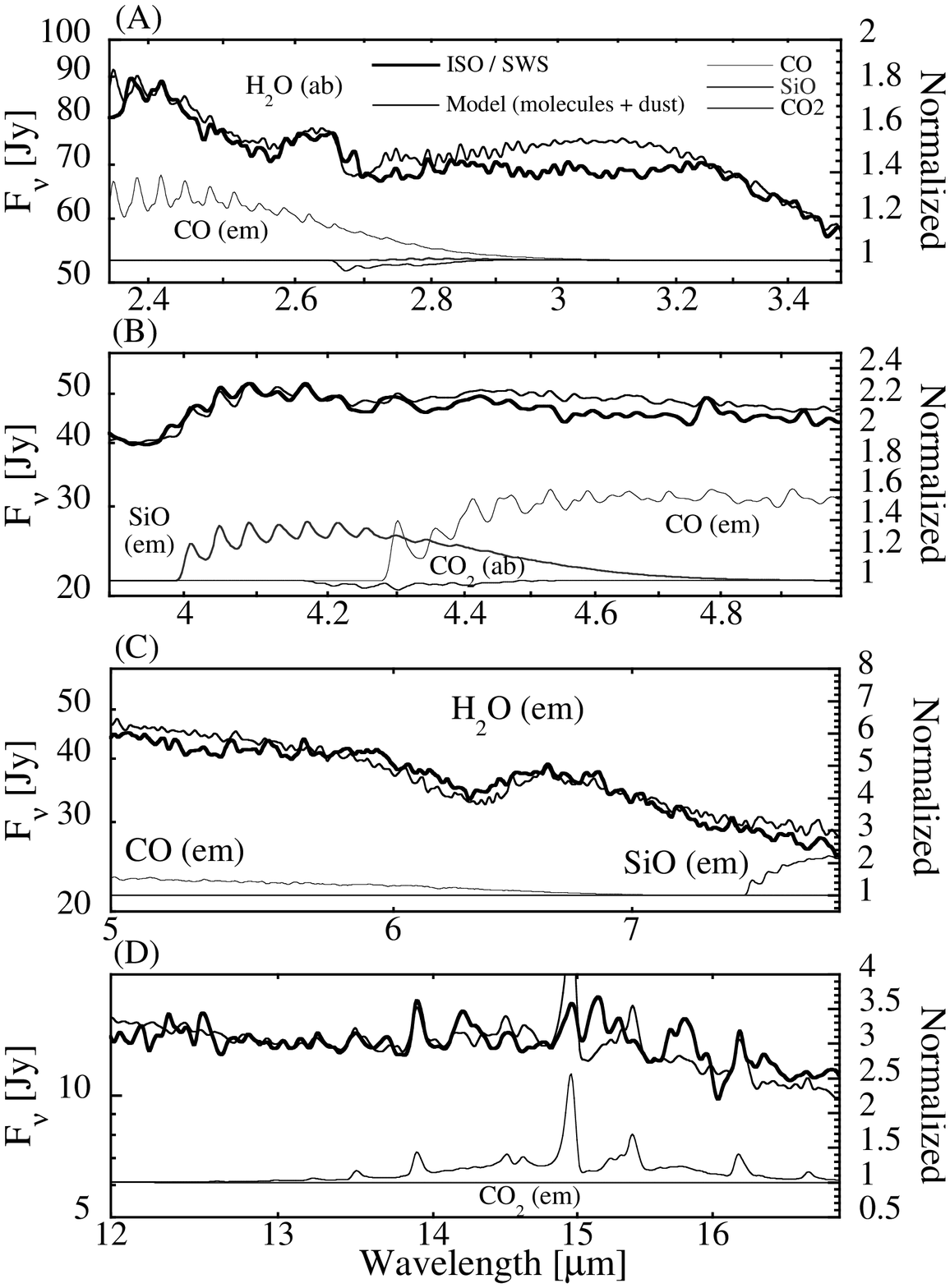}}
\caption{
The ISO/SWS spectra of R~Sct  compared with the synthesized spectra
(the left scale).
Each molecular component is normalized to the background source
and is indicated on the right scale.
The quality of the ISO/SWS data is lower above 13~$\mu$m
and the line intensity of each CO$_2$ line is not reliable.
}
    \label{mol_spec}
\end{figure}

The ISO spectra of R~Sct are compared with the synthesized spectra in
Fig.~\ref{rsct_model}(A) (see also Fig.~\ref{mol_spec}).  The spectra
are well fitted with the parameters listed in
Table~\ref{table-parameters}.
The global shape of the spectra are due to H$_2$O
(Fig.~\ref{rsct_model}(B)).  To demonstrate that the water features
are actually in emission, we also plot separately the background
source (a 5000~K black body).  The H$_2$O features can be fitted with
two components (H$_2$O-1, H$_2$O-2): one component gives the emission
while the other component contributes the absorption at 2.7~$\mu$m
(but also produces weak emission features above 6~$\mu$m). 

In Fig.~\ref{mol_spec}, each molecular component of a synthesized
spectra (CO, SiO and CO$_2$) are compared with the ISO spectra.  By
comparing the observed and synthesized spectra, the observed features
can be identified with H$_2$O, SiO, CO, and CO$_2$ bands either in
absorption or in emission.  SiO and CO$_2$ are identified in R~Sct for
the first time. 
We assumed the same excitation temperature of SiO as
H$_2$O-1 in the synthesized spectra. A weak
absorption feature near 3.1~$\mu$m cannot be explained by H$_2$O 
(Fig.~\ref{mol_spec}.(A)) and is also not
found in $o$~Cet.  This feature can be attributed to either HCN or
C$_2$H$_2$.  HCN is tentatively detected in the millimeter wavelengths
(Bujarrabal et al.\ \cite{Bujarrabal88}) and it could be a candidate
for the 3.1~$\mu$m feature.

Radiation from the dust grains was included in the synthesized
spectra. The dust shell was assumed to surround the molecular extended
atmosphere; the dust has a $r^{-2}$ density distribution as would be expected
for a constant mass-loss rate.  The temperature of the dust grains was
calculated by solving the energy equilibrium, assuming the optically thin
condition. For the calculation of the dust temperature distribution,
the central star was assumed to be a black body of 5000\,K.  The
star was then replaced by the molecular model spectra, and the radiation
transfer was re-calculated.  We used the dust opacity of `warm
oxygen-deficient circumstellar dust' (set 1) in Ossenkopf et al.\
(\cite{Ossenkopf92}).

The parameters required to reproduce the 10~$\mu$m silicate emission
are a dust mass-loss rate of 
$\dot M_{\mathrm d}=1.5 \times 10^{-11}$\,M$_{\sun}$\,yr$^{-1}$, 
and an inner radius of 50\,$R_*$,
assuming the outflow velocity is 10\,km\,s$^{-1}$ (Bujarrabal et al.\
\cite{Bujarrabal88}). The stellar radius ($R_*$) was taken as
$6\times10^{12}$\,cm, and the luminosity as 4000\,L$_{\sun}$.  The outer
radius of the dust shell in Fig.~\ref{rsct_model} is 
$ R_{\mathrm o} =5\times10^4$\,$R_*$. 
However, the spectra can be well reproduced within the
range of $R_{\mathrm o} =10^3$--$10^5$\,$R_*$, and we do not well
constrain $R_{\mathrm o}$.  Note that the radii of the molecular layers
are measured relative to the stellar radius; the assumed value of
$R_*$=$6\times10^{12}$\,cm is only used for the dust model.

\section{Discussion}

\subsection{Evidence for molecules in the extended atmosphere}

Several molecules are detected in the ISO/SWS spectra of R~Sct.  The
range of excitation temperatures (800--4000~K) is lower than that
of the underlying star, confirming that these
molecules are not photospheric.  Several kinds of molecules, with
different transition energies, are detected in emission.  The
implication is that these molecules extend beyond the background
continuum source (the star).

In thermal equilibrium, theoretical model atmospheres predict that
H$_2$O and CO$_2$ are stable in atmospheres cooler than 2000--2500~K
(i.e., late M) and that SiO is stable below 3000~K
(Tsuji~\cite{Tsuji64}).  The effective temperature of R~Sct is
4750--5250~K (Shenton et al.\ \cite{Shenton94}).  Thus, these three
molecules must be located in `an extended atmosphere', a region cooler
than the photosphere of R Sct, i.e. located above the photosphere.

For infrared CO emission, non-LTE effects have been proposed
(summarized in Oudmaijer et al.\ \cite{Oudmaijer95}).  However, these
non-LTE effects have
difficulty explaining all the emission features, since several molecules
with various transition energy are seen in emission at the same time.
It is most likely that the extension of the atmosphere around the star is
responsible for the emission bands.

R~Sct occasionally shows TiO absorption in the visible spectra at
minimum.
TiO is present in the photospheres of cool stars, and
is used for the spectral classification of M-type stars.  
Its presence has been used to assign R Sct a spectral class 
as late as M4 occasionally at minimum phase
(e.g. Preston et al.\ \cite{Preston63};
Cardelli~\cite{Cardelli89}). 
The phase of the ISO/SWS observation is 0.60
(i.e. intermediate between deep minima).  It is unlikely that the
spectral type was early M at the moment of the ISO/SWS observation,
and there is no evidence that the star ever shows a spectral type of
late M required for stable presence of water-vapour. 
This supports the conclusion that
these molecules are not photospheric.
It seems unlikely that stellar pulsation on its own can reduce the stellar
temperature from $\sim 5000\,$K to the $3500\,$K required for the
formation of TiO. Even this molecule may be located
in the extended atmosphere. 

The spectral type of R~Sct, which is later than that of other RV~Tau
stars, may reflect in part its extended atmosphere.  The spectral
energy distribution is enhanced in the infrared region, due to the
presence of the non-photospheric molecular emission bands in infrared.
The resultant effective temperature would underestimate the actual
temperature of the photosphere.  Without the extended atmosphere, the
effective temperature of R~Sct could be more similar to that of other
RV~Tau stars.

\subsection{The envelope structure of R~Sct}

Although the molecules are not photospheric, the excitation
temperatures (800--4000\,K) and radii (1--10\,$R_*$) show that these
molecules are still located near the star.  This suggests that they
formed recently. 

In our SWS data, the absorption features of H$_2$O are seen superimposed
on the H$_2$O emission features. The cooler layer causing the
absorption features and the warmer layer causing the emission are
therefore seen along the same line on sight, with the cooler component
masking the spectra from the warmer layer.  This is possible for a
disk structure if it is seen approximately edge-on, but in general
such a configuration is easier to explain if the atmosphere consists
of an extended shell with a more spherical shape.

The extended atmosphere proposed here describes the
region above but still close to the photosphere, located inside the dust
formation radius, at a temperature of $T \sim 1000\,\rm K$. The
region outside the dust formation radius is referred to as the
circumstellar envelope. Extended molecular atmospheres are seen in
pulsating Mira variables, where they are caused by the pulsations
(e.g., Scholz \cite{Scholz01}).
The extended atmosphere bridges the region between the photosphere and
the dust forming region, and supplies gas from the photosphere to the
dust forming region.  The extended
atmosphere and the circumstellar envelope can be distinguished by the
temperatures and the velocity profiles of the molecules. The molecules
in the extended atmosphere show complicated velocity profiles, often
varying with a time scale of one period, and sometimes moving inward.
The circumstellar envelope traces the dust-driven wind, where the
velocity can be approximated with a near-constant expansion velocity.

There have been previous indications of this extended atmosphere
in R~Sct.
Mozurkewich et al.\ (\cite{Mozurkewich87}) found evidence for velocity
variations with a period of 142 days in the molecular 
absorption lines.
The variation is attributed to 
the pulsation shocks in the extended atmosphere.
This variation is also found in sodium line
(L\`{e}bre \& Gillet \cite{Lebre91}).  
Hinkle et al.\ (\cite{Hinkle96}) reported that CO
showed complicated velocity profiles, while OH and H$_2$O, which have
lower dissociation energies, showed single velocity profiles. All
molecular lines showed time variability.  They concluded that CO is located
near the photosphere and is strongly affected by the stellar
pulsation, while H$_2$O is present in a more extended atmosphere.

The pulsation shocks propagating into the inner region of the
molecular atmosphere should not be so strong that the molecules are
completely destroyed.  However, the shocks can initially aid the
formation of some molecules (Duari et al.\ \cite{Duari99}).  The
molecules may subsequently accumulate in a region above the sodium
layer.

 We used multiple shells as an approximation to the structure of the
atmosphere.  In reality, a single shell with a temperature and density
gradient is more likely, where the inner, warmer region contributes to
the emission and the outer, cooler region causes the absorption on the
emission bands.  The temperature and the density will show
discontinuities at the pulsation shocks, which could still lead to
multiple discrete shells.  However, the shocks do not reach 10~$R_*$,
and the outer shell is expected to be more continuous.

Recently, H$_2$O molecules have been found in K- and M-giants and
supergiants whose effective temperatures are too high for the presence
of water (Tsuji~\cite{Tsuji97}, \cite{Tsuji01}; Jennings \&
Sada~\cite{Jennings98}).  These H$_2$O molecules are located 
above atmospheres, as we have found for R~Sct.  
The extra atmosphere with molecules is commonly seen
in cool red (super-)giants.

Many RV Tau stars show an under-abundance of refractory elements in
the photospheric spectra, similar to depletion patterns seen in the
interstellar medium.  To explain this effect, re-accretion of gas from
a long-lived binary disk has been proposed where gas and dust have
become separated (Waters et al.\,\cite{Waters92}).
Such a
disk has been reported for the RV~Tau star AC~Her 
(Van~Winckel et al.\,\cite{VanWinckel98}; Jura et al.\,\cite{Jura00}). 
These disks are
located in the circumstellar envelope.
R Sct shows a relatively small depletion (Giridhar et al.\
\cite{Giridhar00}). Giridhar et al.\ argued that RV~Tau stars with
late spectral type, such as R~Sct, have enlarged atmospheres: the
accreted, depleted gas mixes with stellar, non-depleted gas in the
extended atmosphere and reduces the depletion.  For R Sct, there is
strong evidence that the enlarged atmosphere exists from the
ISO/SWS observations.

\subsection{Variability}

In Table~\ref{table-ir}, we summarize published infrared photometric
data of R~Sct.  One set of observations, taken near visible minimum
($\phi$=0.07, JD=24\,433\,051), shows fainter magnitudes at all
infrared bands.  Except for this event, the magnitudes are comparable
at all phases.  The magnitudes estimated from the ISO/SWS spectra are
also consistent with other observations at other phases.  
At the $N$-band,
H$_2$O emission increases the flux by a factor of two compared
to the brightness of the central star (Fig.~\ref{ir_spec}). The
constancy of the infrared magnitudes suggests that the molecules were
continuously present around the star over the 30 years of observation,
and did not form shortly before the time of the ISO/SWS observations.

The visual light curve of R Sct shows a number of episodes of deep
minima over this time, one coinciding with the ISO observations and
another with the IRAS observations (Fig.~\ref{fig-lightcurve}). Before
this time, the activity was generally less. The extended molecular
layers may be related to this increased activity, but no spectroscopic
or infrared photometric data is available covering the earlier, less
active period.  With an expansion velocity of 5\,km\,s$^{-1}$, the
hot water layer lags behind the pulsation of the star from the
photosphere
by less than a year.
It is therefore of interest that R~Sct has been relatively
quiescent over the past 3 years: the effect of this on the water layer
could be investigated.

\subsection{Mass-loss history}\label{mass-loss}

\begin{figure}
   \resizebox{\hsize}{!}{\includegraphics*[80,260][520,560]{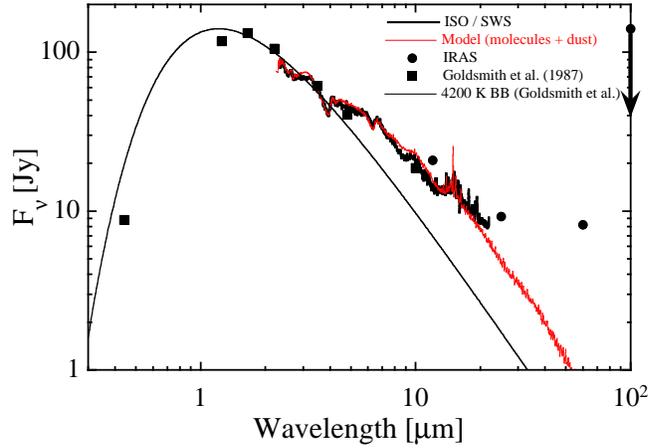}
}
\caption{The ISO/SWS spectra are shown with the IRAS fluxes (the
100~$\mu$m band is an upper limit) and UV and IR photometric data
(Goldsmith et al.\ \cite{Goldsmith87}).  A 4200~K black body, fitting
the UV and IR photometric data (Goldsmith et al.\ \cite{Goldsmith87})
is also plotted.  Zero points and effective wavelength for the filters
from Evans (\cite{Evans93}) are adopted for the data in Goldsmith et
al.\ (\cite{Goldsmith87}) } \label{ir_spec}
\end{figure}

R~Sct shows an excess in the IRAS 60~$\mu$m band (Fig.~\ref{ir_spec}).
There is also a weak 12~$\mu$m excess in the IRAS band, but our
results show that this is mostly due to the H$_2$O emission in the
10~$\mu$m region, with only a minor contribution from dust
(Fig.~\ref{rsct_model}).  The dust mass-loss rate which we estimate
from the 10~$\mu$m silicate emission is 
$ \dot M_{\mathrm d} = 1.5\times 10^{-11}\rm\, M_{\sun}\,yr^{-1}$. 
This mass-loss rate cannot produce the observed
60~$\mu$m excess, and an additional cold dust component is needed.

Alcolea \& Bujarrabal (\cite{Alcolea91}) adopt a two-component dust
shell model based on the 0.4--100~$\mu$m spectra.  For the inner
shell they derive a dust mass-loss rate of 
$\dot M_{\mathrm d}= 4.2\times 10^{-11}\rm \,M_{\sun}\,yr^{-1}$, 
which is comparable to our result
from the 10~$\mu$m excess.  For the outer shell they derive
$\dot M_{\mathrm d}= 2.4 \times 10^{-9}\rm\,M_{\sun}\,yr^{-1}$.  
The CO-derived gas mass-loss rate from
$^{12}$CO $J$=1--0 and 2--1 lines is 
$\dot M = 2\times 10^{-7} \rm\,M_{\sun}\,yr^{-1}$ 
(Bujarrabal et al.\ \cite{Bujarrabal88}).  If a
gas-to-dust ratio of 100 is assumed, the gas mass-loss rate derived
from CO is comparable to the mass-loss rate of the outer shell estimated
from the dust shell model.  Also, the low excitation temperature of
the millimeter CO rotation line suggests that the CO traces the
past mass-loss rate.  The implication is that the mass-loss rate of R
Sct has reduced by about a factor of 100 (Alcolea \&
Bujarrabal~\cite{Alcolea91}).  
Using the inner radius of the outer (cold)
dust shell of $3.7\times10^{16}$\,cm and the CO expansion velocity
of 5~km\,s$^{-1}$, Alcolea \& Bujarrabal estimated the elapsed time
after the large mass loss ceased as $\sim 2000\,\rm yr$.  This
issue will be discussed in the next section.

The low current dust mass-loss rate raises a peculiar point.  The SiO
column density derived here is comparable to that of Mira variables
(Yamamura et al.\ \cite{Yamamura99}).  R~Sct should therefore have a
sufficiently large amount of SiO for silicate dust formation.
However, the current dust mass-loss rate is two orders of magnitudes
lower than that of Mira variables 
(typically 
$\dot M_{\mathrm d}=10^{-9}$~M$_{\sun}$\,yr$^{-1}$--$10^{-8}$~M$_{\sun}$\,yr$^{-1}$).
This suggest a process to inhibit the dust formation from the
molecules. 
The hotter radiation field from the star, compared to
Miras, may affect the dust formation sequence.

For AGB stars, the extended atmosphere is thought to be built up by
pulsations, while radiation pressure on dust is the main
mechanism driving the mass loss (e.g., Fleischer et
al.\ \cite{Fleischer92}; H\"ofner \& Dorfi~\cite{Hoefner97}).  It is
therefore conceivable that both low metallicity and less 
dust grains
operate at small mass-loss in R Sct, even if the intermediate reservoir of SiO has the
same size as in Mira variables.

\subsection{Evolutionary status}\label{evolution}

 RV~Tau stars are commonly assumed to be post-AGB stars with low
initial mass.  For the RV Tau stars with IRAS 60~$\mu$m excess, 
the cessation of the mass loss occurred about five hundreds
years ago (Jura~\cite{Jura86}).  R~Sct also shows evidence for a
larger mass-loss rate in the past, and is also suggested to be in the
post-AGB phase (Alcolea \& Bujarrabal~\cite{Alcolea91}).

The mass loss ceased at the time the star would have left the AGB.
The effective temperature ($T_{\mathrm{eff}}$) at AGB is in the range
of 2000--3000~K, and the current effective temperature of R Sct is in
the range 4000--6000~K. Using the time since the mass-loss reduction
of 2000 yr, we obtain a rate of increase of the stellar temperature
${\mathrm d}T_{\mathrm{eff}}/{\mathrm d}t = 1$--$1.5\rm \, K\, yr^{-1}$.  This
rate can be compared to calculations by Bl\"ocker (\cite{Bloecker95})
of the temperature gradient as function of the post-AGB evolutionary
time scale (in terms of $\log {\mathrm d }t / {\mathrm d}
T_{\mathrm{eff}}$\,yr\,K$^{-1}$, Fig.~6 in his paper).  The post-AGB
time is calculated from the end of the high mass-loss rate.  The rate
we find for R Sct is reproduced for the early post-AGB evolution by
the model tracks of stellar (core) mass range of 0.61, 0.57 and
0.55~M$_{\sun}$. These masses are at the lower end of post-AGB mass
ranges, and may indicate a progenitor star with low initial mass.

Together with the evolution towards higher $T_{\mathrm{eff}}$, the period
of the star will also reduce as the stellar radius becomes smaller. R
Sct is one of the oldest variable stars known, discovered in 1785 by
Edward Pigott, observing from York. He observed the star regularly
until 1803.  Later it was observed extensively by Argelander from
Bonn, between 1843 and 1862, and very good coverage is available from
this time. This long-term light curve allows us to search for evidence
of period evolution.

If all RV Tau stars are post-AGB, they are expected to evolve at
constant luminosity towards higher effective temperatures. We use the
period equation for AGB stars
\begin{equation}
 \log P = 1.949 \log R -0.9 \log M - 2.07
\end{equation}
from Wood (\cite{Wood90}), 
where $P$ is the period in days,
$R$ and $M$ are the stellar radius and mass 
in solar units. Taking
$  L = 4 \pi R^2 T_{\mathrm{eff}}^4 = {\mathrm {const}} $,
where $L$ is the luminosity,
we can rewrite the equation as

\begin{equation}
 \log P = -3.898 \log T_{\mathrm{eff}} +0.975 \log L -0.9 \log M - 2.62.
\end{equation}

This implies that the rate of period change is closely
related to the temperature evolution:

\begin{equation}\label{dotP}
\frac{\dot P}{P} \approx -3.9 \frac{\dot T_{\mathrm{eff}}}{T_{\mathrm{eff}}}
\end{equation}

The right-hand side corresponds to the calculated temperature increase
of $\sim 1\rm \, K\, yr^{-1}$. Using the present period of 140 days, a
change of $- 11$ days per century is expected. This should be easily
detectable in the amateur archives and historical observations.  Zsoldos
(\cite{Zsoldos95}) has noted the lack of observed secular period changes in
other RV~Tau stars and stressed that the low-mass post-AGB interpretation
must be considered uncertain.

\begin{figure*}
\resizebox{\hsize}{!}{\includegraphics*[42,483][528,587]{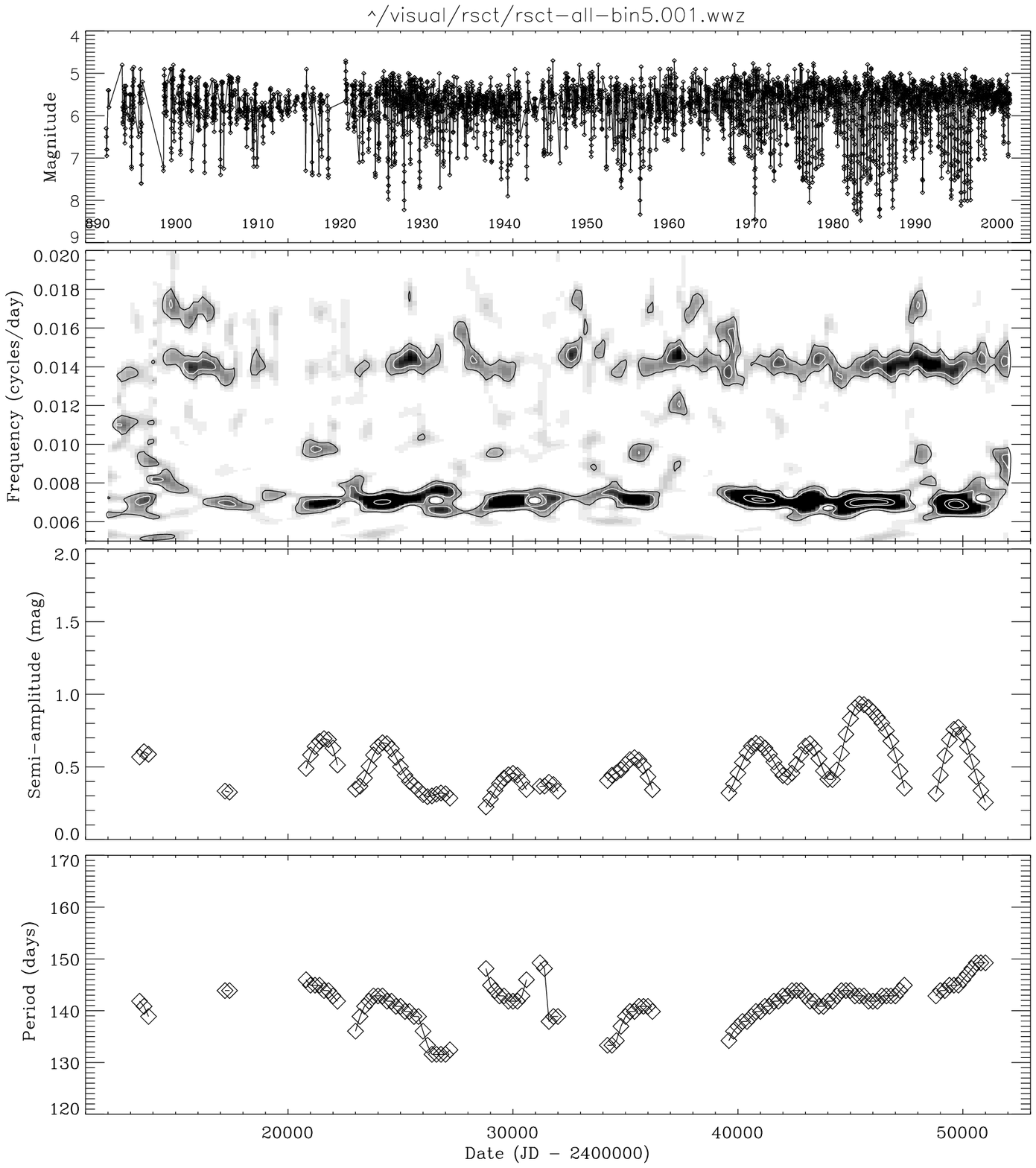}}
\resizebox{\hsize}{!}{\includegraphics*[42,68][528,198]{H3336F61.ps}}
\resizebox{\hsize}{!}{\includegraphics*[42,58][528,198]{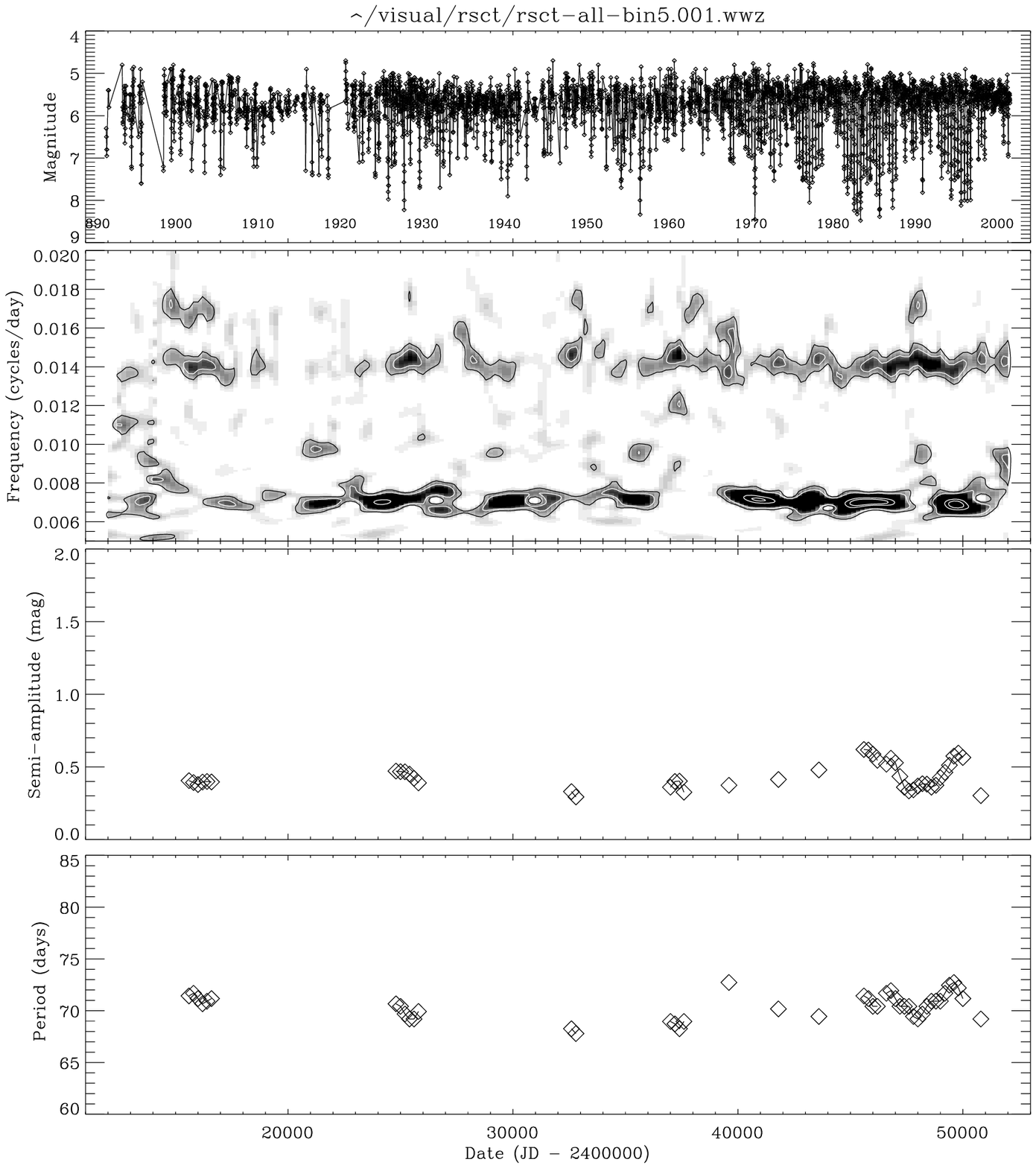}}
\caption{The visual light curve over the 100 years (top) and its wavelet
analysis (the second panel is for the primary period and the third panel is
for the secondary period).  The horizontal scale is $\mbox{JD} - 2400000$.
} \label{fig-lightcurve}
\end{figure*}

We have analyzed published observations of R~Sct to search for
evidence of period decrease.  Individual visual observations were
taken from the databases of the
AFOEV\footnote{http://cdsweb.u-strasbg.fr/afoev/english.htx}, the
BAAVSS\footnote{http://www.britastro.org/vss} and the
VSOLJ\footnote{http://www.kusastro.kyoto-u.ac.jp/vsnet/VSOLJ/vsolj.html},
which extend back over more than a century.  The top panel of
Fig.~\ref{fig-lightcurve} shows the light curve after averaging into
5-day bins.  The deep minima occur with a period of about 140\,days,
and they alternate with the much shallower secondary minima.  
We used
wavelet analysis (Bedding et al.\,\cite{Bedding98}) to measure period
evolution in both the 140-day period and its 70-day harmonic.  These
are shown in the middle and bottom panels of
Fig.~\ref{fig-lightcurve}, respectively.  
While R~Sct shows
substantial period jitter, there is no evidence for a secular trend in
the mean period during the past century.

\begin{figure}
  \resizebox{\hsize}{!}{\includegraphics*[20,420][570,765]{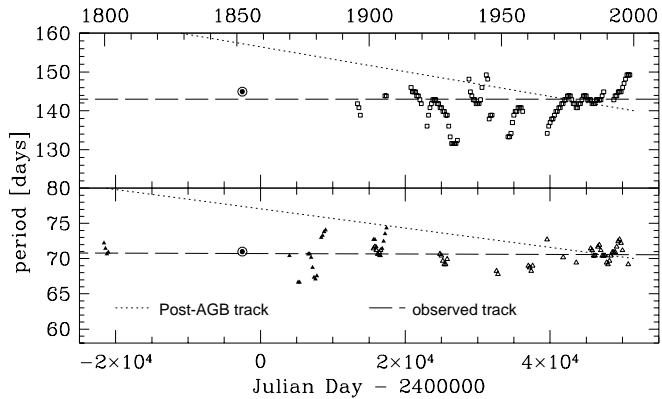}}
\caption{ The period variation of R~Sct over 200 years.  The upper
figure is for the primary period and the lower figure for the harmonic
period.  The upper horizontal scale shows the years.  The predicted
period variation from the post-AGB evolutionary track is shown by the
dotted lines (Sec.\,\ref{evolution}). The observed variation is shown
by the dashed line, where the harmonic period is fitted with the
linear function of $P = [70.71 \pm 0.33] - [0.03 \pm
0.10]\times10^{-4}\times({\rm JD} - 240\,000)$: i.e., the period is
constant over 200~years (long dash line).  }\label{period}
\end{figure}

The evolution of the primary and harmonic periods is plotted in
Fig.~\ref{period} as open symbols.  We also show closed symbols which are
derived from wavelet analysis of earlier published dates of minima.  
Those data are sparse, with some measurements around 1800 and then a large
gap until 1860. 

Period determinations during the 1800s were also made by the observers
themselves.  The values are somewhat scattered, but fall into two
groups: around 60~days and around 70~days. A summary of these
determinations can be found in M\"uller \&\ Hartwig
(\cite{Mueller18}).  As they discuss, Pigott found periods of 63 and
61.5 days, based on the pre-1800 observations. Westphal found 68.8 and
60.56 days, both determinations based on the same data! Argelander
found 60 days in 1844, based on one year of observations, but later
derived 71 days from a 15-year light curve. Schmidt, in 1879, was the
first to derive the fundamental period and found 140 days.

All determinations before Schmidt clearly refer to the harmonic
period.  The reason is that the observers measured the times of maxima
and minima and used these in their analysis: the primary and secondary
minima were not distinguished in the published observations.  Since
the two corresponding frequency groups differ by about 1 cycle per
year, it seems probably that one of these periods is an alias of the
other.  The current (harmonic) period is about 70~days (e.g., 70.6
days measured by Pollard et al.\ \cite{Pollard96}).  It therefore
seems that the (harmonic) period has been about 70~days for as long as
it has been measured.  (Ascribing reality to the 60-day period would
further worsen the discrepancy with eq.~(\ref{dotP}), since it would
give a period evolution with the wrong sign.)

We re-reduced Argelanders observations (\cite{A1869}) to check his
period determination.  The Fourier power spectrum shows a significant
peak at 145.0 days, plus numerous weaker peaks in two ranges, centred
at 140 and 70 days.  Thus we confirm his period, but it is clear that
the true dominant period was already evident in these data. This
analysis also confirms the period stability of R Sct.

In conclusion, over the past two hundred years there is no evidence
for a change of more than 2 days in the harmonic period, corresponding
to 4 days in the main period.  This limit is a factor of 5 below the
predictions from Eq.~(\ref{dotP}).

Percy et al.\ (\cite{Percy91}) analyzed 150 years of data for
R~Sct and reported that a sudden decrease of 1.0\,d occurred in the harmonic
period in 1941.  However, their calculation of 0.07\,day/year as the implied
average rate of period change during the 150\,years of observation is
incorrect.  The correct value is ten times smaller ($1/150 = 0.007$\,day/year).
Thus, their conclusion that R~Sct shows a period decrease in good agreement
with evolutionary models is not valid.  

The lack of evidence for period evolution implies that R~Sct is presently
not evolving towards higher temperature.  This is in contrast to the
post-AGB tracks and suggests alternative evolutionary phases should be
explored, in agreement with the discussion by Zsoldos (\cite{Zsoldos95}).

The past mass-loss rate of R~Sct is of order
$10^{-7}$~M$_{\sun}$\,yr$^{-1}$.  This is rather low for the final
mass-loss rate of the AGB, which is expected to be 
more than
$\sim 10^{-5}$~M$_{\sun}$\,yr$^{-1}$ (Bowen \& Willson~\cite{Bowen91}). 
In addition, the abundance of $s$-process elements for R~Sct is low.
One possible explanation is that
R Sct may not yet have reached the tip of the AGB
(Giridhar et al.\ \cite{Giridhar00}).

It is therefore possible that R~Sct is still an AGB star.
A sudden decrease of the mass-loss rate on the AGB is expected
following the thermal pulse: the lower luminosity during the quiescent
helium burning gives both a higher temperature and a lower mass-loss
rate (Zijlstra et al.\ \cite{Zijlstra92}, Vassiliadis \&
Wood~\cite{Vassiliadis93}, Bl\"ocker\cite{Bloecker95}). The period
evolution slows down and reverses during the helium-burning phase.
The phase lasts for $\sim 5000\,\rm yr$. This phase may fit the period
evolution, and time scale for the detached dust shell, quite well.

We therefore suggest the possibility that R Sct is not a post-AGB
star, but is \object{an AGB star} in the helium-burning phase of the
thermal-pulse cycle.

\section{Conclusions}

We find H$_2$O, SiO, CO$_2$, and CO in the ISO/SWS spectra of R~Sct.
H$_2$O emission dominates the near- and mid-infrared spectra.  The
molecules are probably located in a spherical extended
atmosphere.  Pulsations may gradually lift up the gas from the
photosphere to the extended atmosphere.  Molecular formation appears
to take place in the extended atmosphere.

The column density estimated from the infrared SiO bands is comparable
to that in Mira variables, however, the current dust
mass-loss rate of
R~Sct is lower than Mira variables by two orders of magnitudes.  The
high effective temperature may prevent dust formation around this
star.

While a high mass-loss rate in the past is indicated by an excess at
60~$\mu$m, and by the CO millimeter wavelength emission, the current
mass-loss rate is two orders of magnitude lower: the dust mass-loss
rate $\dot M_{\mathrm d} = 10^{-11}$\,M$_{\sun}$\,yr$^{-1}$.  The
reduction in the mass-loss rate, and the time scale, can be explained
if R~Sct is a low-mass post-AGB star. However, the period evolution is
too slow for post-AGB evolution. We therefore suggest that R Sct may
be an AGB star in the helium-burning phase of the thermal-pulse cycle.

As future work, monitoring of the infrared photometric bands could
help in understanding the structure of the extended atmosphere 
in this star, as well as in Mira variables.  The infrared light curve of Mira
variables shows a phase delay of maximum phase by about $\phi$=0.1. For
the infrared variability of Mira variables, dust might be important
(e.g. Winters et al.\ \cite{Winters94}), but molecules in the pulsating
atmosphere contribute as well (Matsuura et al.\ \cite{Matsuura02}).  If
the same infrared phase delay is found for R~Sct
the contribution of molecules
on the infrared light-curve will be strengthened.  In
Table~\ref{table-ir}, the infrared photometry shows slightly fainter
magnitudes at phase $\phi=$0.07 (JD=244,3051.31).  More frequent
observations of the infrared light curve could find a phase delay.

\begin{acknowledgements}
We acknowledge Dr. Mattei and the American Association of Variable
Star Observers (AAVSO) for light curve of R~Sct.  Ten-micron data for
R~Sct, listed in the appendix, was taken with TIMMI-2 on ESO 3.6 m
telescope.  We gratefully thank the instrumental team and support
astronomers for their assistance.  The advice by Dr. Kester for
IRAS was useful.  This research was supported by PPARC grant.
I.Y. acknowledges support by Grant-in-Aid for Encouragement of
Young Scientists (No. 13740131) from Japan Society for the Promotion
of Science.
\end{acknowledgements}

\appendix
\section{Infrared photometry}

  Infrared photometric data taken from the literature are listed in
Table~\ref{table-ir}.  The data are taken from Gehrz (\cite{Gehrz72}),
Lloyd Evans (\cite{LloydEvans85}), Goldsmith et
al.\ (\cite{Goldsmith87}), and Shenton et al.\ (\cite{Shenton94}).  The
zero points for the IRAS bands are taken from Wainscoat et al.
(\cite{Wainscoat92}).  The flux density of the ISO/SWS spectra is
converted to magnitude, using filter transmission in van der Bliek et
al.\ (\cite{vanderBliek96}) and zero points in Evans (\cite{Evans93})
the for L- and M-bands. For N- and Q-bands,
effective centre wavelengths and widths in Evans (\cite{Evans93})
are used.
We measured an N-band
magnitude with TIMMI-2 on 3.6\,m Telescope of ESO on 4 August, 2001,
using a narrow band filter at 10.4\,$\mu$m.  HD\,133774 was used as a
calibration source adopting the template in Cohen (\cite{Cohen99}).
The zero point for the 10.4\,$\mu$m filter is calculated from the
spectrum of Vega (Cohen et al.\ \cite{Cohen95}).

\begin{table*}
\begin{caption}
{
Infrared photometric data
}
\label{table-ir}
\end{caption}
\begin{tabular}{r@{.}lc cc r@{.}l r@{.}l  r@{.}l r@{.}l r@{.}l c cc c} \hline
\multicolumn{2}{c}{JD} & Phase & $J$ & $H$ & \multicolumn{2}{c}{$K$} &
  \multicolumn{2}{c}{$L$} & \multicolumn{2}{c}{$M$} &
  \multicolumn{2}{c}{$N$} & \multicolumn{2}{c}{$Q$} & IRAS~12 & IRAS~25 & Ref 
\\
\multicolumn{2}{c}{2440000+}&      &      &\multicolumn{2}{c}{}&
  \multicolumn{2}{c}{}&\multicolumn{2}{c}{}&\multicolumn{2}{c}{}&
  \multicolumn{2}{c}{}&  \\ \hline
\multicolumn{2}{r}{Uncertain$^{\ddag}$}        &      &      &      &
  2&8  & 1&8  & 1&4  & 0&9$^{\ddag}$&\multicolumn{2}{c}{}&&& 1  \\
2682&31 & 0.64 & 2.88 & 2.33 & 2&04 & 1&53 &\multicolumn{2}{c}{}&
  \multicolumn{2}{c}{}&\multicolumn{2}{c}{}&&& 2 \\
2999&36 & 0.72 & 2.78 & 2.28 & 2&02 & 1&51 &\multicolumn{2}{c}{}&
  \multicolumn{2}{c}{}&\multicolumn{2}{c}{}&&& 2 \\
3051&31 & 0.07 & 3.38 & 3.04 & 2&80 & 2&33 &\multicolumn{2}{c}{}&
  \multicolumn{2}{c}{}&\multicolumn{2}{c}{}&&& 2 \\
6243&61 & 0.61 & 2.84 & 2.26 & 2&00 & 1&49 &\multicolumn{2}{c}{}&
  \multicolumn{2}{c}{}&\multicolumn{2}{c}{}&&& 3 \\
6248&53 & 0.64 &      &      &\multicolumn{2}{c}{}&\multicolumn{2}{c}{}&
  1&50 & 0&85 &\multicolumn{2}{c}{}&&& 3 \\
6253&50 & 0.68 & 2.84 & 2.26 & 2&00 & 1&49 &\multicolumn{2}{c}{}&
  \multicolumn{2}{c}{}&\multicolumn{2}{c}{}&&& 3 \\
6991&48 & 0.80 & 3.10 & 2.52 & 2&26 & 1&77 & 1&15 & 0&36 & 0&46 &&& 4 \\
7000&49 & 0.86 & 3.10 & 2.49 & 2&22 & 1&65 &\multicolumn{2}{c}{}&
  \multicolumn{2}{c}{}&\multicolumn{2}{c}{}&&& 4 \\
7040&25 & 0.15 & 3.11 & 2.57 & 2&36 & 1&91 & 1&67 & 0&93 & $-$0&01 &&& 4 \\
8840&36 & 0.49 & 3.06 & 2.46 & 2&26 & 1&91 &\multicolumn{2}{c}{}&
  \multicolumn{2}{c}{}&\multicolumn{2}{c}{}&&& 4 \\
8895&36 & 0.86 & 2.84 & 2.31 & 2&10 & 1&76 &\multicolumn{2}{c}{}&
  \multicolumn{2}{c}{}&\multicolumn{2}{c}{}&&& 4 \\
  \multicolumn{2}{r}{Uncertain}&      &      & &
  \multicolumn{2}{c}{}& \multicolumn{2}{c}{}&\multicolumn{2}{c}{}&
  \multicolumn{2}{c}{}&\multicolumn{2}{c}{}& 0.74$^{\S}$ & 0.03$^{\S}$ & 5 \\
10153&24 & 0.60 &     &      &\multicolumn{2}{c}{}& 1&81$^{\P}$ &
  1&36$^{\P}$ & 0&98$^{\P}$ & 0&00$^{\P}$     & 0.75$^{\S}$ & 0.05$^{\S}$ & 6 
\\
12125&55 &     &      &      &\multicolumn{2}{c}{}
  &\multicolumn{2}{c}{}&
  \multicolumn{2}{c}{} & \multicolumn{2}{c}{0.75} &\multicolumn{2}{c}{}&&& 7 
\\ \hline
\end{tabular}\\
1: Gehrz (\cite{Gehrz72}).
2: Lloyd Evans (\cite{LloydEvans85}).
3: Goldsmith et al.\ (\cite{Goldsmith87}).
4: Shenton et al.\ (\cite{Shenton94}).
5: IRAS data. IRAS observed R~Sct three times, and
    no variability has been found (Var flag is 0~\%).
6: Estimate from the ISO/SWS data. Filter transmission of van der Bliek et al.
  ($L$, $M$; \cite{vanderBliek96}) and
  IRAS Explanatory Supplement (\cite{iras-es}) are adopted.
  $N$ and $Q$ band are estimated from effective centre wavelength and width in
  Evans (\cite{Evans93}).
7: ESO/TIMMI-2 observations. Not broad $N$-broad band but narrow $N$-band 
filter
  ($N$10.4).\\
$^{\dag}$ Observed between April 1970 and April 1971.
$^{\ddag}$ Using a narrow band filter centred at 10.8~$\mu$m.
$^{\P}$ Zero magnitudes are taken from Evans (\cite{Evans93}).
$^{\S}$ Zero magnitudes are taken from Wainscoat et al.\ 
(\cite{Wainscoat92}).\\
Phases from JD=2442682.31 to JD=2446253.50 are estimated from
the light curve in Mattei (\cite{Mattei85}; \cite{Mattei90}).
\end{table*}

\end{document}